\input{aipcheck}
\documentclass[numberedheadings]{aipproc}
\layoutstyle{6x9}

\usepackage{epsfig}

\begin{document}
\title{\bf News on Leptogenesis
 \footnote{Proceedings of the Third Tropical Workshop, 
{\em Neutrinos, Branes and Cosmology}, San Juan, Puerto Rico, 19-23 August, 2002.
Based on the reference papers \cite{bdp},\cite{bdp2}.}}

\author{P. Di Bari}{address={ 
{\it\small IFAE, Universitat Aut{\`o}noma de Barcelona,  08193 Bellaterra (Barcelona), Spain} \\
{\it\small e-mail: dibari@ifae.es}}}
\date{}


\begin{abstract}
The possibility to explain the CMB measurement of the 
baryon asymmetry with leptogenesis results in  a stringent 
bound on neutrino masses such that $\sqrt{m^2_1+m^2_2+m^2_3}< 0.30\,{\rm eV}$. 
We discuss the implications of such a bound for future experiments  
on the absolute neutrino mass scale. 
\end{abstract}

\maketitle

\section{The baryon asymmetry of the Universe}

The observation of the acoustic peaks in the power spectrum of CMB 
temperature anysotropies confirms that we live in a baryon asymmetric 
Universe, an important result already inferred from the study of 
cosmic rays and primordial nuclear abundances. 
Within standard BBN (SBBN) any measurement of a primordial nuclear abundance 
leads to a measurement of the baryon asymmetry, 
conveniently expressed in the form of baryon to photon number ratio. 
 From the measurement of the Deuterium primordial abundance in
Quasar absorption systems one finds at $1\sigma$ \cite{Omeara}:
\begin{equation}
\left.\eta_{B0}^{SBBN}\right|_{D/H}=(5.6\pm 0.5)\times 10^{-10}
\end{equation}
A multiple measurement of different primordial abundances represents a test of consistency
for SBBN and in principle should lead to a more accurate 
determination of $\eta_{B0}$. However the results from the
Helium and Litium primordial abundances are only marginally consistent
with the Deuterium abundance and thus it is necessary to 
account for larger sistematic uncertainties and to make some assumptions
on their statistical distribution. Thus an acceptable 
agreement among the abundances leads to a less precise determination of 
the baryon asymmetry \cite{Fields}
\begin{equation}\label{SBBN}
\eta_{B0}^{SBBN}=(2.6-6.2)\times 10^{-10},
\end{equation}
valid approximately at the $90\%$ c.l. \cite{private}.
The difficulty of SBBN in explaining simultaneously all
the current measurements of primordial abundances can also be interpreted
as a hint for the presence of non standard BBN effects.
 Some of them are well motivated within those models beyond the 
Standard Model that can incorporate the see-saw and lead to leptogenesis. 
In any case a determination of the baryon asymmetry 
from SBBN, at a higher 
level of accuracy than the range (\ref{SBBN}), 
encounters serious obstacles at the present.

Fortunately the recent observation of acoustic peaks in the power spectrum of CMB
temperature anysotropies provides a powerful tool to measure the baryon asymmetry
and to circumvent the difficulties of SBBN. In this case one has a 
good consistency of different determinations of the baryon asymmetry 
from 6 different
experiments employing different tecniques \cite{CMB,Archeops}. 
A recent combined analysis gives \cite{Archeops}
\begin{equation}\label{etaCMB}
\eta_{B0}^{CMB}=(6.0^{+ 0.8}_{-1.1})\times 10^{-10}.
\end{equation}
This determination is in reasonable agreement with that one from 
the SBBN and at the same level of accuracy. However, in contrast with
the SBBN determination, the consistency of the different experimental results 
so far makes it quite robust and makes possible to expect a reduction of the error 
in a close future: below the $10\%$ level from the MAP satellite during next years
and at the 1\% level from the Planck satellite before the end of this decade.
 We will therefore use the CMB determination of the baryon asymmetry in 
our following considerations.

\section{Basics of leptogenesis}

Leptogenesis \cite{fy86} is the cosmological consequence  
of the see-saw mechanism. This explains the lightness of neutrino
masses by the existence of three RH neutrinos, $N_{i}$, much heavier than
the electroweak scale. The decay of the heavy neutrinos violates  
lepton number and, in general, also CP conservation,  while the cosmological expansion  
can yield the necessary departure from thermal equilibrium: all three
Sacharov's conditions are satisfied and a lepton number can be generated in the early 
Universe. The possibility for leptogenesis to explain the observed baryon asymmetry
relies crucially on the existence of the non perturbative SM sphaleron processes, 
that can convert, at temperatures above the electroweak phase transition,
 about $-1/3$ of the  lepton number  into a baryon number, while keeping B-L constant. 
The source of CP violation is naturally provided by the complexity of the
neutrino mass matrices in the see-saw. For each of the three $N_i$
one can introduce a CP asymmetry parameter defined as:
\begin{equation}
\varepsilon_i \equiv {\Gamma_i-\bar{\Gamma}_i\over \Gamma_i +\bar{\Gamma}_i},
\end{equation}
where $\Gamma_i$ and $\bar{\Gamma}_i$ are the decay rates of
$N_i$ respectively into leptons ($N_i\rightarrow l+\bar{\phi}$) 
and anti-leptons ($N_i\rightarrow \bar{l}+\phi$).

The problem is greatly simplified if one assumes that 
only the decays of the lightest RH neutrinos, $N_1$,
can influence the final baryon asymmetry.  This is true 
if the asymmetries generated by the two heavier neutrino decays (with masses 
$M_2$ and $M_3$), even though  not negligible, are subsequently washed out by the processes
(for example inverse decays) in which the lightest right-handed neutrinos 
(with mass $M_1$) are involved, at temperatures $T \sim M_1$. 
This assumption implies the existence of a mild hierarchy of masses such that 
$M_{2,3}\stackrel{>}{\sim} (2-3)\,M_1$ and also that the
wash out $N_1$-processes are strong enough. 

In this way one has to solve a system of only two Boltzmann equations, 
one for the number of $N_1$'s and one for the $B-L$ asymmetry. 
Introducing the convenient variable $z\equiv M_1/T$,
they can be written in the following simple form \cite{Luty,Plumacher,bdp}:
\begin{eqnarray}\label{ke}
{dN_{N_1}\over dz} & = & -(D+S)\,(N_{N_1}-N_{N_1}^{\rm eq}) \;, \label{lg1} \\ \label{ke2}
{dN_{B-L}\over dz} & = & 
-\varepsilon_1\,D\,(N_{N_1}-N_{N_1}^{\rm eq})-W\,N_{B-L} \;.\label{lg2}
\end{eqnarray}
There are four classes of processes that contribute to the
different terms in the equations: 
decays, inverse decays, $\Delta L=1$ scatterings and RH neutrino mediated
processes. The first three contribute all together to modify the
$N_1$ abundance. Indicating with $H$ the expansion rate,
the term $D\equiv \Gamma_D/(H\,z)$ accounts for the
decays and inverse decays while the term $S\equiv\Gamma_S/(H\,z)$
accounts for the $\Delta L=1$ scatterings.
The decays are also the source term for the generation of the $B-L$ asymmetry,
the first term in the second equation, while all the other processes 
contribute to the wash out term $W\equiv \Gamma_W/(H\,z)$ that competes 
with the decay source term.

\section{A model independent parameterization}

From the Eq. (\ref{ke2}) it is easy to see that 
the solution $N_{B-L}(z)$ has to depend linearly on $\varepsilon_1$,
in a way that the final baryon asymmetry can be written in the form
\begin{equation}
N_{B-L}^{\rm fin}=N_{B-L}^{\rm in}-{3\over 4}\varepsilon_1\,\kappa_0\,\, .
\end{equation}
Assuming that the wash out processes are strong enough to erase 
an initial value of $N_{B-L}$, generated for example by the decays of the two heavier
RH neutrinos or by some other unspecified mechanism, we will put 
$N_{B-L}^{\rm in}=0$. This assumption is valid under the same conditions 
for which heavier neutrino decays can be neglected and therefore it does not
introduce further restrictions.

The {\em efficiency factor} $\kappa_0$ does not depend on $\varepsilon_1$.
It is normalized in a way to be $1$ in the limit case that an initial thermal 
abundance of $N_1$'s decays fully out of equilbrium at the time when all wash out 
processes are completely frozen. In this limit the wash out term in the 
kinetic equations is uneffective and can be neglected. 
Let us introduce the quantity:
\begin{equation}\label{barm}
\bar{m}=\sqrt{m_1^2+m_2^2+m_3^2}
\end{equation} 
The {\em quadratic mean of the light neutrino masses} is simply 
$\bar{m}/\sqrt{3}$.  A remarkable fact is that for masses 
$M_1\ll 10^{14}\,{\rm GeV}\,(0.1\,{\rm eV})/\bar{m})^2$ the three terms $D,S$ and $W$
are proportional to an {\em effective neutrino mass} $\widetilde{m}_1$
times a function of $z$ alone \cite{Plumacher}. This means that the final baryon asymmetry, 
for small $M_1$, will depend only on two parameters: $\varepsilon_1$
and $\widetilde{m}_1$. In this case
the {\em out of equilibrium limit} is obtained 
for $\widetilde{m}_1\rightarrow 0$.
In figure 1 we show the function $\kappa_0$ as a function of $\widetilde{m}_1$,
for different values of $M_1$. It can be seen how for small values
of $M_1$ there is no dependence on $M_1$ itself. We performed the calculations both
for an initial thermal abundance (thin lines) and for a zero initial abundance
(thick lines). It is evident how there is a critical value of 
$\widetilde{m}_1$ that separates two different regimes.
For $\widetilde{m}_1 \ll 5\times 10^{-4}\,{\rm eV}$ one recovers the limit
of out of equilibrium decays and $\kappa_0$ is strongly dependent on the number
of initial $N_1$'s. In the case of zero initial neutrinos, $\kappa_0$ 
is determined by the number of $N_1$'s that
are produced by inverse decays and scatterings and this number goes to zero
in the limit $\widetilde{m}_1\rightarrow 0$. Therefore in this regime  
there is a strong dependence on the initial conditions.
For $\widetilde{m}_1\gg 5\times 10^{-4}\,{\rm eV}$ there is no dependence on the initial
conditions and, even for  a zero initial number of $N_1$'s, they are rapidly produced 
and their number rapidly approaches the thermal value. This means that, in the limit 
of large values of $\widetilde{m}_1$, 
the dependence of $\kappa_0$ on the $N_1$ production processes disappears  
and only a dependence on the wash out processes is left.
 
In the intermediate regime the value of $\kappa_0$ is determined by an interplay 
between the wash out processes and the number of decaying $N_1's$, 
determined both by the initial number and by the strenght of production processes. 
It is possible to give a numerical fit of $\kappa_0$:
\begin{equation}
\kappa_0=f_-(x_-)\,e^{-x_-}+f_+(x_+)\,e^{-x_+} \;,
\end{equation}
with
$x_{\pm}=\left(\widetilde{m}_1 / \widetilde{m}_{\pm} \right)^{\alpha_{\pm}}$.
The first term depends on the initial $N_1$ abundance: for an initial zero abundance
\begin{equation}
f_-(x_-)=f_+(x_-)=0.24\,x_- ,\;\;\widetilde{m}_{-} = 3.5 \times 10^{-4}\,{\rm eV},
\;\;\alpha_-=0.9\, ;
\end{equation} 
 for an initial thermal abundance
\begin{equation}
f_-(x_-)=1 ,\;\;  \widetilde{m}_{-} = 4.0 \times 10^{-4}\,{\rm eV},\;\;
\alpha_-=0.7\, .
 \end{equation}
The second term is independent on the initial conditions:
\begin{equation}
f_+(x_+)=0.24\,x_+,\;\;\;\widetilde{m}_{+} = 8.3 \times 10^{-4}\,{\rm eV},\;\;\; \alpha_{+} = -1.1.
\end{equation}

\begin{figure}
\hspace{5mm}
\psfig{figure=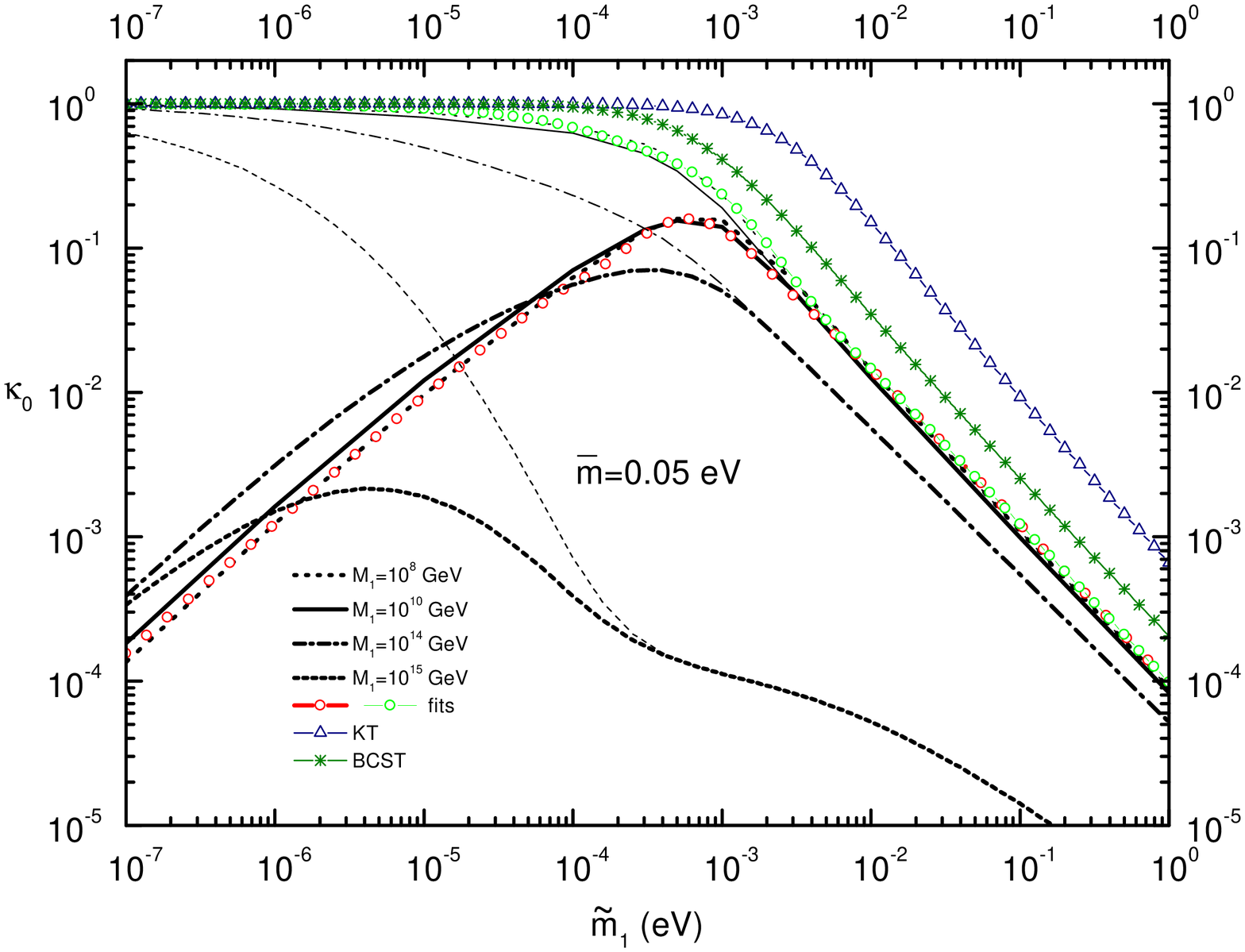,height=9cm,width=14cm}
\caption{The efficiency factor.}
\end{figure}

 In the limit of weak coupling ($\widetilde{m}_1\ll 5\times 10^{-4}\,{\rm eV}$) one has
$\kappa_0\simeq f_-(x_-)$, while in the limit
of strong coupling ($\widetilde{m}_1\gg 5\times 10^{-4}\,{\rm eV}$) one has:
\begin{equation}\label{kappa0}
\kappa_0\simeq f_+(x_+)\simeq 10^{-4}\,
\left({\rm eV}\over \widetilde{m}_1\right)^{1.1}\, .
\end{equation}
The two fits are optimal for $M_1=10^{8}\,{\rm GeV}$ and
are represented in figure 1 with circled lines.
It is interesting to compare these results with an analytical approximation
for $\kappa_0$ originally derived in the context of GUT baryogenesis \cite{kt} but
also adapted to the case of leptogenesis (see for example \cite{yasu}):
\begin{eqnarray}
\kappa_0 & = & 1\;\;\;\; {\mbox {\rm for}}\,\,\; K\ll 1 \\
\kappa_0 & =  & {0.3\over K\,(\ln K)^{0.6}} \;\;\;\; {\mbox{\rm for}}\,\,\; K\gg 1
\end{eqnarray}
As a simple interpolating expression we can use:
\begin{equation}\label{kt}
\kappa_0={0.3\over K\,[\ln(1+K)]^{0.6}}\,\left( 1+{0.3\over K\,[\ln (1+K)]^{0.6}}\right)^{-1}
\end{equation}
The quantity $K$ is related to $\widetilde{m}_1$ simply by:
\begin{equation}
K= {1\over 2}\,D|_{z=1}\simeq 170\,{\widetilde{m}_1\over {\rm eV}}
\end{equation}
The expression (\ref{kt}) is represented in the figure 1 with the triangle line. 
One can see that it overestimates the efficiency factor by $\sim 7$. This is 
not surprising because this analytical approximation takes into account
only the inverse decays in the wash out term and it neglects
the other processes that are equivalently important.  
A more specific analytical approach was described in \cite{barbieri}
and the result for $\kappa_0(\widetilde{m}_1)$ is represented in the figure 1 with 
the starry line 
\footnote{We deduced it from the figure 1 in \cite{barbieri} interpolating the 
points for $M_1=10^{8}\,{\rm GeV}$. Strangely the result does not 
correspond to the analytical expression that is given in the text
and numbered as Eq. (4.3).}. One can see that it better 
agrees with the numerical results but it still overestimates 
them by a factor $2-3$.

For large values of $M_1$ the efficiency factor depends also 
on $M_1$ itself and actually for $M_1\stackrel{>}{\sim} 10^{14}\,{\rm GeV}\,
(0.1\,{\rm eV})/\bar{m})^2$ there 
is a suppression in the regime for large $\widetilde{m}_1$. 
The suppression is due to a term 
$\Delta W \propto M_1\,\bar{m}^2/z^2$, originating
from the RH neutrinos mediated processes, that, for large $M_1$, 
dominates in the total wash-out term $W$ \cite{bdp}.  
This term suppresses exponentially the baryon asymmetry yielding a  
term $\exp{(-{\rm const}\,M_1\,\bar{m}^2/\bar{z})}$ in the efficiency factor,
where $\bar{z}$ is that value of $z$, larger than $1$, at which $\Delta W$ starts to dominate and it can depend only on $M_1$, $\bar{m}$ and $\widetilde{m}_1$. 
Thus , in the most general case,
the final baryon asymmetry can be described in terms of only four parameters: $\varepsilon_1$,
$\widetilde{m}_1$, $M_1$ and $\bar{m}$. 
 
\section{The surface of maximum baryon asymmetry}

In order to obtain a prediction for
$\eta_{B0}$, to be compared with the measured value $\eta_{B0}^{CMB}$
in the Eq.(\ref{etaCMB}), one has to multiply $N_{B-L}^{\rm fin}$  
for the fraction of the $B-L$ asymmetry that is converted into baryons by the sphaleron processes, given by a factor $28/79\simeq 1/3$, and divide for the {\em dilution factor} $f=N_{\gamma}^{\star}/N_{\gamma}^{0}$. This takes into account that the generated baryon asymmetry gets diluted compared to the number
of photons that are produced in the annihilations of all standard model particle species. If one assumes a standard thermal history of the early Universe then $f\simeq 28$ and in the end one gets the simple relation
\begin{equation}
\eta_{B0}\simeq -10^{-2}\,\varepsilon_1\,\kappa_0\, .
\end{equation}
A first trivial model independent bound on $\eta_{B0}$ is obtained 
considering that  $|\varepsilon_1|\leq 1$ and thus 
$\eta_{B0}\stackrel{<}{\sim} 10^{-2}\,\kappa_0$. It is however possible
to show a more stringent bound on $|\varepsilon_1|$ 
\cite{cpbound,barbieri,di,bdp,bdp2}:
\begin{equation}\label{CPbound}
|\varepsilon_1|< 
{3\over 16\pi}\ {M_1\over {\rm v}^2}\ {m^2_3-m^2_1 \over m_3}\;
= 10^{-6}\,\left({M_1\over 10^{10}\,{\rm GeV}}\right)\,\beta
\end{equation}
From neutrino mixing experiments  
$m^2_3-m^2_1=\Delta m^2_{\rm atm}+\Delta m^2_{\rm sol}$ and 
one can write: 
\begin{equation}\label{beta}
\beta \simeq 
{\Delta m^2_{\rm atm}+\Delta m^2_{\rm sol}\over 0.051\,{\rm eV}\,m_3}
\end{equation}
For example in the case of hierarchical neutrinos, for $m_1=0$, one has  
{\small $m_3= \sqrt{\Delta m^2_{\rm atm}+\Delta m^2_{\rm sol}}$}
and $\beta \simeq 1$.
While in the case of quasi degenerate neutrinos with 
$\bar{m}\simeq 1\,{\rm eV}$, one has 
$m_3\simeq 0.58\,{\rm eV}$ and $\beta\simeq 0.1$.
In general $m_3$, in the Eq. (\ref{beta}), has to be regarded as
a function of $\bar{m}$. This is true if one considers 
the quantities $m^2_3-m^2_2$ and $m^2_2-m^2_1$ fixed 
by the solar and atmospheric neutrino experiments. There are
two possibilities. In the case of {\em normal hierarchy} 
\begin{eqnarray} \label{nD32}
m^2_3-m^2_2 & = &  \Delta m^2_{\rm atm}\, , \\
m^2_2-m^2_1 & = &  \Delta m^2_{\rm sol} \label{nD21} 
\end{eqnarray} 
and inverting the Eq.'s (\ref{barm}), (\ref{nD32}) and (\ref{nD21}) one finds:
\begin{eqnarray}
m_3^2 &=& {1\over 3}\left(\bar{m}^2 + 2\Delta m^2_{\rm atm} + \Delta m^2_{\rm sol}\right)\;, \\
m_2^2 &=& {1\over 3}\left(\bar{m}^2 - \Delta m^2_{\rm atm} + \Delta m^2_{\rm sol}\right)\;, \\
m_1^2 &=& {1\over 3}\left(\bar{m}^2 - \Delta m^2_{\rm atm} - 2\Delta m^2_{\rm sol}\right)\;. 
\end{eqnarray}
These relations are plotted in figure 2. 
\begin{figure}
\hspace{10mm}
\psfig{figure=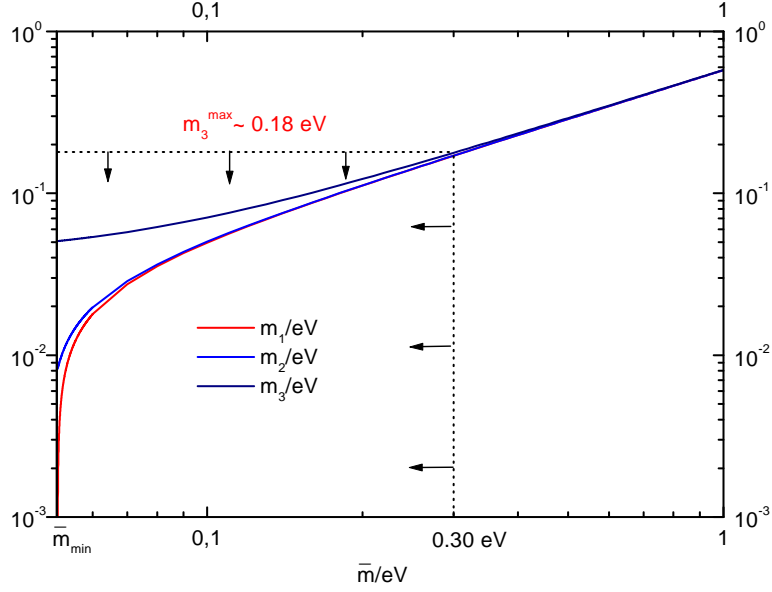,height=90mm,width=13cm}
\caption{Dependence of the light neutrino masses on $\bar{m}$  
in the case of normal hierarchy.}
\end{figure}
In the case
of {\em inverted hierarchy} 
\begin{eqnarray} \label{iD32}
m^2_3-m^2_2 & = &  \Delta m^2_{\rm sol}\, , \\
m^2_2-m^2_1 & = &  \Delta m^2_{\rm atm} \label{iD21} 
\end{eqnarray} 
and inverting the Eq.'s (\ref{barm}), (\ref{iD32}) and (\ref{iD21}) one finds:
\begin{eqnarray}
m_3^2 &=& {1\over 3}\left(\bar{m}^2 + \Delta m^2_{\rm atm} + 2\,\Delta m^2_{\rm sol}\right)\;, \\
m_2^2 &=& {1\over 3}\left(\bar{m}^2 + \Delta m^2_{\rm atm} - \Delta m^2_{\rm sol}\right)\;, \\
m_1^2 &=& {1\over 3}\left(\bar{m}^2 - 2\,\Delta m^2_{\rm atm} - \Delta m^2_{\rm sol}\right)\;. 
\end{eqnarray}
Using the relation $m_3(\bar{m})$ in the Eq.(\ref{beta}) 
one immediately gets the general dependence of  $\beta$ on $\bar{m}$.

From the CP bound, Eq. (\ref{CPbound}), one can see that, given the atmospheric neutrino mass scale, the mass of the lightest RH neutrino cannot be higher than $10^{16}\,{\rm GeV}$. This
because otherwise $|\varepsilon_1|$ would be, absurdly, higher than $1$.  This has to be
also consistently derived within the see-saw formula. It is certainly true 
in the oversimplified case of one generation see-saw formula: 
if $m\stackrel{>}{\sim} 0.05\,{\rm eV}$ then $M\stackrel{<}{\sim} 10^{15}\,{\rm GeV}$. 
For three generations the result is analogous 
and again consistent with the CP bound (see for example \cite{willen})
as it has to be. Thus in the end one can
express the maximum baryon asymmetry in terms
of just three parameters $\widetilde{m}_1$, $M_1$ and $\bar{m}$:
\begin{equation}\label{etamax}
\eta_{B0}^{\rm max} \simeq 10^{-8}\,\beta(\bar{m})\left({M_1\over 10^{10}\,{\rm GeV}}\right)\,\kappa_0(\widetilde{m}_1,M_1,\bar{m})
\end{equation} 
This is the surface of maximum baryon asymmetry and the CMB constraint
is given by the requirement that $\eta_{B0}^{\rm max}\geq \eta_{B0}^{CMB}$.
\begin{figure}
\hspace{5mm}
\psfig{figure=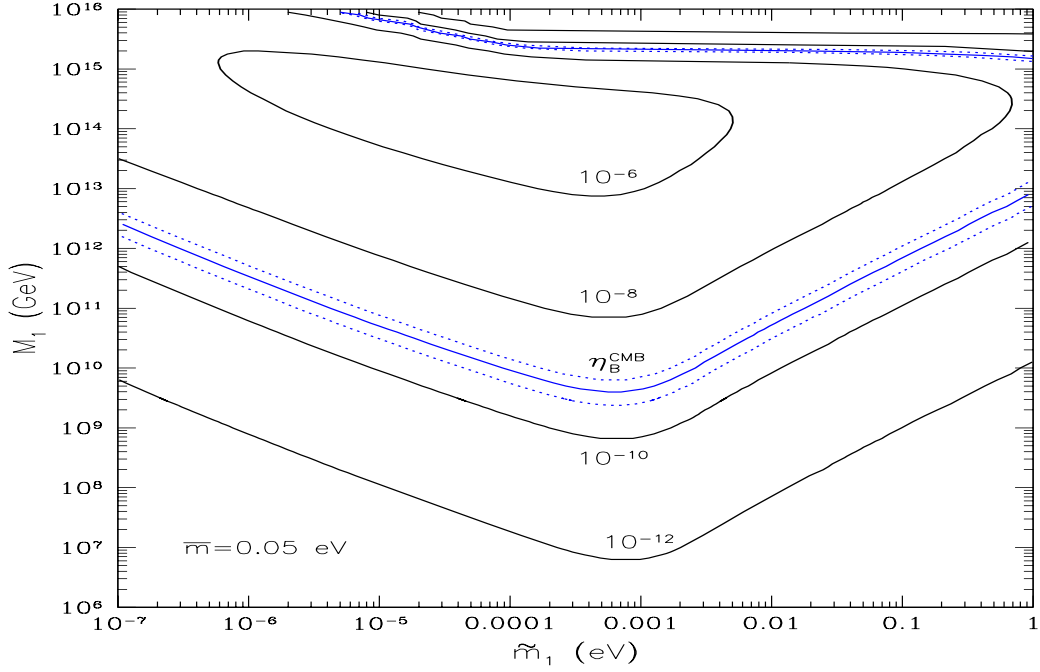,height=9cm,width=14cm}
\caption{Lines of constant $\eta_{B0}^{\rm max}$ in the plane 
$(\widetilde{m}_1,M_1)$ (from \cite{bdp}).}
\end{figure}
In figure 3 we show the iso-$\eta_{B0}^{\rm max}$ lines in the plane $(\widetilde{m}_1,M_1)$
in the case of a zero initial abundance of $N_1$'s and
for $\bar{m}$ equal to its minimum value that is obtained 
in the case of normal hierarchy:
\begin{equation}
\bar{m}_{\rm min}\equiv \sqrt{\Delta m^{2}_{\rm atm}+ 2\,\Delta m^2_{\rm sol}}\simeq 0.05\,{\rm eV},
\end{equation}
This value implies $\beta\simeq 1$ in the bound (\ref{CPbound}) 
on $|\epsilon_1|$ and also $m_1=0$. The {\em allowed region} lies within the contour line
for $\eta_{B0}^{\rm max}=(\eta_{B0}^{CMB})_{\rm low}$ (the external dotted line).
 There is clearly a lower bound on $M_1$. This can be directly
obtained from the Eq. (\ref{etamax}) imposing $\kappa_0$ and $\beta\leq 1$ 
\cite{bdp}:
\begin{equation} 
T_L\simeq M_1\geq 10^{8}\,{\rm GeV}\,{\eta_{B0}^{CMB}\over 10^{-10}}
\stackrel{>}{\sim} 4\times 10^{8}\,{\rm GeV},
\end{equation}
at $\sim 2\,\sigma$ from the Eq. (\ref{etaCMB}). In the case of a zero initial 
abundance one can see from the figure 1 that $\kappa_0\leq 0.16$ and thus a 
more stringent constraint, as visible in figure 3, follows:
\begin{equation}
T_L\simeq M_1\geq 6.25\times 10^{8}\,{\rm GeV}\,{\eta_{B0}^{CMB}\over 10^{-10}}\stackrel{>}{\sim} 2.5 \times 10^{9}\,{\rm GeV}.
\end{equation}

\section{Bound on neutrino masses}

If $\bar{m}$ increases then the allowed region shrinks. This happens
both because the bound on the CP asymmetry $|\varepsilon_1|$ gets more restrictive
($\beta$ gets smaller in the Eq. (\ref{CPbound}) ) and because the action of the
wash-out term $\Delta W\propto M_1\,\bar{m}^2$ gets stronger. 
At the same time if $\bar{m}>\bar{m}_{\rm min}$ then $m_1>0$ and in this
case it is possible to show another important constraint \cite{hamaguchi}:
\begin{equation}
\widetilde{m}_1 \geq m_1
\end{equation}
Increasing $\bar{m}$ there will be a value for which the two constraints
together cannot be simultaneously satisfied. In
figure 4 
\begin{figure}
\hspace{5mm}
\psfig{figure=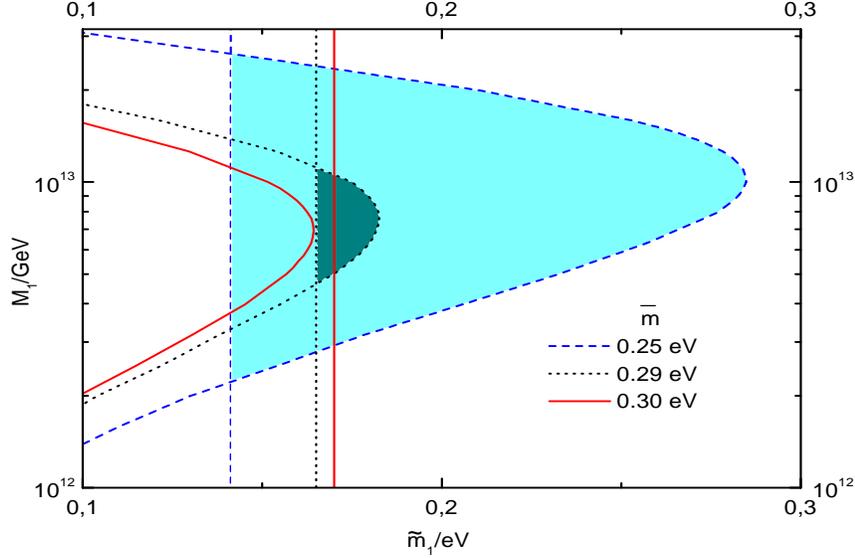,height=85mm,width=14cm}
\caption{Bound on $\bar{m}$ (from \cite{bdp2}). The vertical lines 
correspond to
$\widetilde{m}_1=m_1$. The other curves are the lines corresponding to
$\eta_{B0}^{\rm max}=(\eta_{B0}^{CMB})_{\rm low}$.  
There is no allowed region (the filled areas) for $\bar{m}=0.30\,{\rm eV}$.} 
\end{figure}
it is clearly shown that, with a precision of $0.01\,{\rm eV}$, this value 
is given by  $\bar{m}=0.30\,{\rm eV}$ \cite{bdp2} and this can be considered an upper limit
on $\bar{m}$ for leptogenesis to explain the observed baryon asymmetry.
From figure 2 one can see  that in terms of light neutrino masses this limit 
corresponds to have 
$m_3\stackrel{<}{\sim} 0.18\,{\rm eV}$ and $m_1\simeq m_2 
\stackrel{<}{\sim} 0.17\,{\rm eV}$ 
and this means that 
{\em leptogenesis is incompatible with quasi-degenerate light neutrinos.} 

This conclusion has relevant implications for those future experiments that 
will try to determine the {\em absolute mass scale of neutrinos}. 
There are three classes of such experiments:
tritium $\beta$ decay experiments, neutrinoless double $\beta$ decay experiments
and cosmological experiments. 

The {\em tritium $\beta$ decay experiments} are sensitive to an effective electron
neutrino mass given by:
\begin{equation}
m_{\nu_e}=\sqrt{\sum_{i=1}^{3}\,|U_{ei}|^2\,m^2_{i}}
\end{equation}
where the $U_{e i}$ are the elements of the mixing matrix 
for the electron neutrino flavour. Since $U$ is unitary then $m_{\nu_e}\leq m_3$.
Therefore the leptogenesis bound predicts that $m_{\nu_e}<0.18\,{\rm eV}$.
The KATRIN experiment \cite{KATRIN} will be sensitive to $m_{\nu_e}
\stackrel{>}{\sim} 0.30\,{\rm eV}$
(at $90\%$ c.l.) and thus a compatibility with the leptogenesis bound
implies no detection of a positive signal.

The {\em cosmological experiments} are able to place
stringent limits on the {\em sum of neutrino masses}. In this case the
leptogenesis bound predicts:
\begin{equation}
\sum_i\,m_i < 0.52\,{\rm eV}
\end{equation}
The data from the MAP and Planck satellites combined with those from the 
SLOAN Digital Sky Survey will be sensitive 
(at $1\,\sigma$) to $ \sum_i\,m_i \stackrel{>}{\sim} 0.23\,{\rm eV}$ and
$ \sum_i\,m_i \stackrel{>}{\sim} 0.06\,{\rm eV}$ respectively \cite{Hannestad}.  Therefore there is
certainly room for a positive signal compatible with the leptogenesis bound.

The neutrinoless double $\beta$ decay experiments are sensitive, in the case 
of Majorana neutrino masses as from the see-saw, to the quantity:
\begin{equation}
m_{ee}=\left|\sum_{i=1}^3\,U^2_{ei}\,m_i \right|
\end{equation}
In this case the leptogenesis bound implies:
\begin{equation}
m_{ee}< 0.18\,{\rm eV}
\end{equation}
Future experiments, like the GENIUS project \cite{genius}, should be sensitive to 
$m_{ee}> {\cal O}(0.01)\,{\rm eV}$ and thus can 
find a positive signal compatible with the leptogenesis bound.
Besides these three classes of experiments it is also
worthwhile to mention that in the fortunate case that 
future generations of neutrino
telescopes will detect the intense neutrino fluxes needed
for the Z burst scenario to explain the UHECR anomaly and if this anomaly
will be confirmed, then 
we will have another powerful method to measure the absolute 
scale of neutrino masses, more specifically the highest 
mass eigenvalue $m_3$ \cite{Ringwald}.

\section{Conclusions}

It is remarkable that the leptogenesis predictions of the final baryon 
asymmetry can be expressed in terms of just 4 parameters, in a
model independent way. This result relies on two main assumptions: the existence
of a mild hierarchy in the masses of the RH neutrinos 
($M_{2,3}\stackrel{>}{\sim} (2-3)\,M_1$)
and that the initial temperature can be assumed to be larger than $M_1$.
With this parameterization one can easily describe the requirements for
a succesfull leptogenesis. A quite precise temperature for leptogenesis seems 
to emerge to explain the observed baryon asymmetry. 
The most striking result is that the light
neutrino masses cannot be too larger than the atmospheric neutrino mass scale
$\sim 0.05\,{\rm eV}$, thus ruling out the class of quasi degenerate neutrino models.
This implies a strong, though negative, prediction on the possibility of 
future experiments to detect a sub-eV neutrino mass scale, unless their sensitivity
can be pushed below ${\cal O}(0.1\,{\rm eV})$. If an evidence for an absolute neutrino
mass scale violating the leptogenesis bound will be found then it is possible to
relax the main assumptions with an enhancement of the CP asymmetry 
from a degeneracy of the RH neutrino masses \cite{deg} or with a
non thermal production of RH neutrinos \cite{nt}. In this cases 
however the nice link between the observed
baryon asymmetry and neutrino masses would be lost or controlled by some
additional adjustable parameter. If the future experimental results will agree with
 the leptogenesis bound then the picture will be certainly highly strengthened by the
 tight conspiracy between the observed baryon asymmetry and neutrino masses.
   
\section*{Acknowledgments}
This work was supported by the EU Fifth Framework network ``Supersymmetry
and the Early Universe" (HPRN-CT-2000-00152).
All the presented results have been obtained in collaboration with 
W. Buchm\"{u}ller and M. Pl\"{u}macher and more details can be found in
\cite{bdp}, \cite{bdp2}. 
I wish to thank J.~Pati who stimulated the comparison of the
results on the efficiency factor with those in \cite{kt} and
J. O'Meara and S. Sarkar for clarifications on the baryon asymmetry measurements.
I also wish to thank the following people for their interest in leptogenesis and for
nice discussions during my year in the theory group of DESY, during SUSY02, 
the Third Tropical Workshop and COSMO02 and during my visits in Fermilab, 
University of Delaware, Bartol Research 
Institute and University of Maryland: K.~Abazajian, J.~Beacom, N.F.~Bell, S.~Bludman, 
 A.~Brandenburg, L.~Covi, J.~Formaggio, E.~Lisi, R.~Fleischer, P.H.~Frampton, A.~De Gouvea, 
S.~Huber, B.~Kyae, C.N.~Leung, M.~Luty, G.~McGregor, R.~ Mohapatra, H.B.~Nielsen, J.~Nieves, J.~Pati, A.~Pilaftsis, A.~Ringwald, R.R.~Volkas, Q.~Shafi, Y.~Takanishi, Y.Y.Y.~Wong.


\end{document}